\documentstyle[12pt,aaspp4]{article}

\begin{document}

\title{The First FIRST Gravitationally Lensed Quasar: FBQ~0951+2635\altaffilmark{1}} 
\author{
Paul L. Schechter\altaffilmark{2}, 
Michael D. Gregg\altaffilmark{3}, 
Robert H. Becker\altaffilmark{3,4},
David J. Helfand\altaffilmark{5} and 
Richard~L.~White\altaffilmark{6} }

\authoraddr{MIT Room 6-206, 77 Massachusetts Ave., Cambridge MA 02139}
\altaffiltext{1}{Observations reported in this paper were obtained in part at
the MDM Observatory, a facility no longer jointly operated by the
University of Michigan, Dartmouth College and the Massachusetts
Institute of Technology}
\altaffiltext{2}{Department of Physics, Massachusetts Institute of Technology, Cambridge MA 02139}
\altaffiltext{3}{Institute of Geophysics and Planetary Physics, Lawrence Livermore National Laboratory, Livermore, CA 94450}
\altaffiltext{4}{Department of Physics, University of California, Davis, CA 95616}
\altaffiltext{5}{Department of Astronomy, Columbia University, 538 West 120th Street, New York,
NY 10027}
\altaffiltext{6}{Space Telescope Science Institute, 3700 San Martin Drive, Baltimore MD 21218}

\begin{abstract}

The $V=16.9$ quasar FBQ 0951+2635 at redshift $z=1.24$ appears double
on CCD exposures taken in subarcsecond seeing.  The two objects are
separated by $1\farcs1$ and differ in brightness by 0.9 mag.  VLA
observations show the radio source to be double with the same
separation and position angle.  Spectra taken with the Keck II telescope
show the two components to have nearly identical emission line
spectra, but with somewhat different absorption line systems.
Subtraction of two stellar point spread functions from the pair of
components consistently leaves a residual object.  Depending upon whether
this third object is extended or a point source it may be as much as
1/10 or as little as 1/100 as bright as the brighter QSO component.  The
observations leave no doubt that the 2 brighter objects are
gravitationally lensed images of the same quasar.  The third object
might be either the lensing galaxy or a third image of the quasar, but
both interpretations have serious shortcomings.

\end{abstract}

\keywords{cosmology: gravitational lensing --- quasars, photometry, spectroscopy}

\section{INTRODUCTION}

Quasars that are multiply imaged by gravitational lenses can be used
for a wide range of astrophysical investigations (e.g., Kochanek and
Hewitt 1996), but the number of known systems is as yet so small that
that their potential has only begun to be realized.  We report here
the discovery of a new gravitationally lensed object, FBQ 0951+2635, a
$V=16.9$, 1.7 mJy, $z=1.24$ quasar drawn from the FIRST Bright QSO
Survey (henceforth FBQS; Gregg {\it et al.} 1996).  It has two
components with a separation of $1\farcs1$ and a radio flux ratio of 4:1.

The FBQS offers two advantages as a source of potential gravitational
lenses.  First, since the quasars are bright, with $R<17.5$, they are
more likely to be lensed than quasars from fainter surveys (Turner
{\it et al.} 1984).  Second, since the FBQS quasars are identified
using the FIRST radio survey (Becker, White and Helfand 1995;
henceforth BWH), radio observations can be used both to confirm the
source as a lensed object and to constrain models of the lens.

With these strengths in mind, we undertook to obtain high resolution
optical images of quasars identified in the FBQS with the Hiltner
2.4-m telescope at Michigan-Dartmouth-MIT (MDM) Observatory. A short
red exposure of FBQ 0951+2635 taken in good seeing immediately revealed
two stellar objects, with a separation of roughly 1'' and a flux ratio
of roughly one magnitude.  Followup images obtained in $V$ and $I$ showed
that the two objects were approximately the same color.  Re-examination
of the Lick spectrum used to obtain the FBQS redshift showed no sign
of stellar features.

Early attempts to obtain spectra of the fainter object were frustrated
by poor weather and poor seeing.  But an A-array map obtained with the
NRAO\footnote{The National Radio Astronomy Observatory is operated by
Associated Universities Inc. under a cooperative agreement with the
National Science Foundation.} Very Large Array (VLA) showed that the
radio source was also double, with the same separation and position
angle, and roughly the same flux ratio.  Radio emission from the
fainter component makes it very unlikely that the object is a star -- fewer
than 0.03\% of FIRST radio sources have Galactic stellar counterparts
(Helfand et al. 1997).  Spectra of the two components obtained
subsequently with the Keck II telescope are very similar, but not
identical.

Longer $B$,$V$,$R$ and $I$ exposures obtained in somewhat better
seeing during a subsequent MDM run in 1997 January showed evidence for
another object not quite colinear with the two quasar components.  Though
one is tempted to identify this object with the lensing galaxy, the
relatively low resolution of the present direct images leaves its
character in doubt.

In the sections that follow we describe in greater detail the
observations, reductions and arguments which lead us to conclude that
FBQ 0951+2635 is gravitationally lensed.  We also examine briefly the
consequences of two alternative hypotheses regarding the nature of the
third object: that it is the lensing galaxy and that it is a third
image of the quasar.  Both hypotheses present serious difficulties.

\section{OBSERVATIONS AND REDUCTIONS}

\subsection{Initial Optical Imaging}

FBQ 0951+2635 is identified as a $z=1.24$, $B \sim 16.9$ quasar in a
forthcoming paper (Becker {\it et al.} 1997) presenting the second
list of FIRST quasars.  A spectrum taken at Lick Observatory shows
broad lines of Mg II, C III and C IV.  The flux at 20 cm is $\sim 2$
mJy.

A 180 s $R$ filter exposure taken with the Hiltner 2.4 m telescope at
MDM Observatory on 1996 December 10 in $0\farcs9$ seeing showed a
double object (cf. Figure 1a), with the fainter component at PA
125$^\circ$.  A 180 s $V$ filter exposure and three 180 s $I$ filter
exposures, all with $1\farcs0$ seeing, were obtained on the same
night.  Following bias subtraction and flatfielding, an empirical
point spread function (PSF) was fit simultaneously to the two objects
using a variant of the program DoPHOT (Schechter {\it et al.} 1993)
designed to deal with close point-like and extended objects (Schechter
and Moore 1993).  The magnitude differences between the brighter and
fainter components (henceforth components ``A'' and ``B'') in $V$, $R$
and $I$ were 0.97, 0.95 and 0.90 mag, respectively, and the
separations were 1\farcs09, 1\farcs08 and 1\farcs07 respectively.  Both trends are
in the sense expected if the lensing galaxy, which should be closer to
the fainter component, contaminates it more than the brighter
component.  The following night, 1996 December 11, was photometric.
Images in $B$, $V$, $R$ and $I$ were obtained, and reduced as
described in \S 2.4 below.

\subsection{Followup Radio Mapping}

The system was observed at 3.6 cm with the VLA in the A configuration
for 20 minutes on 1996 December 31.  The data show clear evidence for
duplicity at the same position angle as the optical data (Figure 2).
This effectively rules out the possibility that the second optical
object is a star randomly projected close to the quasar.  Positions
and fluxes obtained from this observation are given in Table 1.  The
larger separation in the radio than in the optical (1\farcs12 as opposed
to 1\farcs08), if significant, can be understood on the hypothesis that
the fainter component is blended with the lensing galaxy.

\subsection{Followup Optical Imaging}

Additional optical observations were obtained on 1997 January 10 and
11, consisting of 600s exposures in each of $B$, $V$, $R$ and $I$,
with an additional $2 \times 500$ s obtained in $I$.  The seeing on
these images ranged from 0.8 to $0\farcs9$.  When reduced as described
above, the separation decreased monotonically going from blue to red.
There was also a similar pattern of residuals in each of the images
(Figure 3), with a single bright spot slightly east of the A
component and north of the B component, and fainter bright rims 
southwest of each component.  The amplitude of the positive residuals is
quite small, of order 1-2\% of the central intensity of component A.  The
negative ``divots'' at the centers of the two components are 3-4\% of the central
intensity.

The most straightforward interpretation of this pattern of residuals
is that there is a third object, roughly midway between the A and B
components, slightly to the NE.  The two-star model accommodates this
third object by bringing the two components closer, giving them a
larger flux (and hence leaving large divots) and displacing the two
components toward the third object, leaving crescent-like residuals on the
opposite side.  This pattern is most obvious on the $R$ and $I$ images,
less so on the $V$ image and yet less so on the $B$ image, which might be
expected if the third object were the lensing galaxy.

An alternate interpretation is that one is simply seeing the
consequences of an imperfect template, due to variations in the PSF
across the CCD.  There are several reasons to think this is not the
case.  In producing Figure 3, we were careful to use a PSF template
which lay within 100 pixels of the quasar.  Were the residuals the
result of template mismatch, one would see the same pattern associated
with both A and B, which seems not to be the case.  Moreover the
residuals of nearby stars do not show the same pattern.  Using other
stars as templates show similar patterns, especially on the $R$ and $I$
images.

We therefore undertook to obtain a fit for two point sources plus an
extended object (henceforth ``C''), which, for the sake of
argument we take to be the lensing galaxy, again following the
approach of Schechter and Moore (1993).  The ``galaxy'' profile is
taken to be a circularly symmetric pseudogaussian, and the positions
of both quasar components were treated as free parameters.

The scatter in the galaxy position in these four fits was $0\farcs04$ in
the North-South direction and $0\farcs10$ in the East-West direction, in
both cases small compared to the separation of the two components.  The
scatters in R.A. and declination for the positions of component B were
15 and 12 mas respectively.  The ratio of the flux in the galaxy to
that in the brighter component A was roughly
10\% in all filters.

The average $\sigma$ of the pseudogaussian, after correction for
seeing, was $0\farcs510$, which corresponds to a FWHM somewhat larger than
the seeing disk.  The average separation between components A and B is
$1\farcs101$, some 20 mas closer than the radio separation.  This may be an
underestimate, since the redder filters give smaller separations than
the bluer, even after allowing for the presence of the lensing galaxy.
However the radio separation and position angle applied to the optical
image give residuals typical of having enforced too large a separation.

While the trend of smaller separation with redder color appears to
be significant, we have adopted, for the present, a straight average
of the positions in the four filters as our best guesses of the
actual positions, and report these in Table~2.  For the purpose of
computing commensurable magnitudes, we fix the relative separations at
these values.  We likewise fix the width of the ``galaxy'', object C,
at the mean position and shape obtained from the 4 filters.  We then
fit the data solving only for the overall position of the system and
the relative fluxes of the three components, the results of which are
also given in Table~2.

It is useful to compare the residuals for the two cases -- without the
galaxy and with the galaxy and quasar components at their mean positions.
These are shown for the $R$ image in Figure 1, along with a low contrast
image which shows the component positions.  Also shown is the image with A
and B subtracted but with the galaxy unsubtracted.

Table 2 shows that the flux ratio of C to A depends surprisingly little
on filter, and if anything, gets bluer in the bluest
filter.  One would expect the lensing galaxy to be considerably redder
than the quasar.  This leads one to suspect that one is seeing quasar
light and not light from the lensing galaxy.  This could be either the
result of a poor PSF or of a third image of the quasar.  We have
already argued that the third object is not an artifact of poor PSF
fitting.  When we try a fit treating the system as three point
sources, we find that the fluxes attributed to the third object are
very much smaller than when it was treated as a galaxy, and that the
colors are no longer so very blue.  Our tentative conclusions are
first, that the galaxy hypothesis is somewhat more likely, and second,
that we are pressing our data to the limits of what it can tell us.

\subsection{Photometric and Astrometric Reduction} 

The data taken on 1996 December 11 included exposures of FBQ 0951+2635
in somewhat poorer seeing and of Landolt (1992) standards in the
fields of PG 0231+051 and Rubin 152.  The standard and quasar exposures
were reduced using routine IRAF reduction programs.  We have solved
for zeropoint and color terms using ``typical'' KPNO extinction
coefficients (Massey et al. 1997).  Since the standards and the quasar were
observed at nearly the same low airmass, the use of typical
coefficients (which we take to be uncertain by 30\%) introduces errors
of at most 0.02 mag for the bright stars in the quasar fields.
Photometry for the brightest stars in the quasar field is given in
Table 2.

Positions were measured for the brighter stars in the quasar field
using the 600s R exposure from January 10, with an empirical PSF used
for fitting.  This yielded an astrometric solution (relative to the
APM astrometry of McMahon and Irwin 1992) with a fractional
uncertainty in the scale of 1/750 and with a comparable uncertainty in
the rotation of the detector.  Positions for these stars, here
relative to the quasar A component, are also presented in Table 2.  The
zeropoint of our astrometric solution differed from the VLA position
in Table 2 by almost 1'', considerably more than is typical for
APM-VLA comparisons.  The APM catalog position for the composite A+B
shows the same difference.

The empirical PSF fitting described in the previous subsection gives
magnitudes for all pointlike objects relative to a template star.
With the magnitudes for the template stars now measured, these can now
be put on the standard system (after correction for color terms).  The
results for the quasar are given in Table 2.  With a little
calculation the pseudogaussian used to model the lensing galaxy can be
integrated and a total flux derived.  When compared to the reference
stars in the field this gives the magnitudes listed in Table~2.  It
should be noted that our use of a fitting function which falls off
more rapidly than a galaxy profile is almost certain to underestimate
the brightness of the lensing galaxy.

\subsection{Spectroscopy}

Spectra were obtained with the Low Resolution Imaging Spectrograph
(LRIS) on the Keck II telescope on Valentine's Day 1997.  The exposure
was 300 seconds in $0\farcs6$ seeing.  The dispersion of $2\farcs44$
\AA /pixel gave a resolution of 11 \AA.  The slit was was aligned with
the two bright components.  Components A and B are clearly resolved
spatially along the slit (Figure 4), for which the image scale is
$0\farcs215$/pixel.  The two components (Figure 4) are quite similar,
both showing Mg II absorption systems at z=0.73 and z=0.89.  But the
emission lines appear to be weaker in the B component than in the A
component. The continuum of the B component appears to drop faster
than that of the A component beyond 7000 \AA, but this may be the
result of having intentionally removed the order separating filter.
Differences in the equivalent widths of emission lines have been seen
in other gravitationally lensed systems (e.g., Wisotzki {\it et al.}
1993) where it is taken as a sign that the continuum is subject to
microlensing.

\section{MODELS AND INTERPRETATION}

\subsection{The Lensing Galaxy Hypothesis}

\subsubsection{Models for the Lens Potential}

Taking the third object to be the lensing galaxy, we have tried
fitting several alternative models to the observed positions of
components A and B relative to C.  Our simplest model, which has the
basic features of an elliptical galaxy or dark matter halo, is the
singular isothermal quadrupole potential (henceforth SIQP; Kochanek
1991).  It produces a flat rotation curve and has equipotentials which
are self-similar and roughly, though not exactly, elliptical, and
corresponds to a density distribution which is likewise self-similar
but not exactly elliptical.

The model has five parameters: a lens strength, $b$ which is roughly
half the separation of the images, a shear coefficient $\gamma$ (which
for modest quadrupoles is a factor of 6 smaller than the ellipticity
of the underlying mass distribution), an orientation for the shear, and
the two angular coordinates of the source with respect to the optical
axis defined by the observer and the lensing galaxy.  With two pair of
coordinates for the quasar images measured with respect to the lensing
galaxy, we need one additional constraint.  We use the flux ratio of
component B to component A. Mao and Schneider (1997) have noted that
flux ratios are susceptible to micro- and milli-lensing and are
therefore less reliable constraints than the positions.

When we fit this simple model to our observations, we find that the
ellipticity of the density is only 0.05, very much rounder than most
elliptical galaxies.  Alternatively, if one takes the shear to be
tidal in origin (and shear is clearly needed because the galaxy is not
colinear with the two quasar images and because the flux ratio is very
different from unity) one finds a dimensionless shear of only 0.016.

In either case, the quadrupole term is smaller than has been measured
in any lens system.  A neighboring galaxy with the same potential as
the lensing galaxy would have to be at a distance more than 15 times
the separation of the two components for its tidal shear to be as small as
0.016.  Yet there is another object, probably a neighboring galaxy,
some 2" east and slightly south of component B in Figures 1b and 1c.  It
is also visible in the $V$ and $I$ images (Figure 3).

The near equidistance of object C from components A and B also implies
that A and B are both very close to the Einstein ring, and therefore
are very highly magnified, with component A magnified by a factor of 30.
But the probability of a given magnification varies roughly inversely
as the magnification.  The low shear and high magnification indicated
by the positions and flux ratio seem very unlikely.

The redder images place the lensing galaxy closer to the fainter
component, as is expected.  Using only the $I$ filter image gives a
larger shear, $\gamma = 0.11$, and more modest magnifications, 5.9 and
2.5.  But one is hard pressed to explain why the bluer images give a
third object closer to A component.  If there were a central
component, C', contaminating what we have taken to be the galaxy, one
would expect it to be on the same side of the lensing galaxy as the
intermediate distance component.  Component C' should be between the
galaxy and B rather than on the other side.

\subsubsection{The SED of the Lensing Galaxy}

For the purpose of predicting the apparent magnitude of the lensing
galaxy, we assume that the lensing potential is a that of a singular
isothermal sphere with associated line of sight velocity dispersion
$\sigma$. If the potential has strength $b$, the
associated velocity dispersion is given by
\begin{equation}
{\sigma^2 \over c^2} = {D_S \over D_{LS}} {b \over 4 \pi} \quad ,
\end {equation}
where and $D_S$ and $D_{LS}$ are, respectively, angular diameter
distances to the source and from the lens to the source (Narayan and
Bartelmann 1997), with $b$ measured in radians.  For FBQ 0951+2635 $b
\sim 0\farcs55$.  Following Keeton {\it et al.} (1997) we take
\begin{equation}
{L \over L^* } = \left({\sigma \over 220 km/sec}\right)^4 \quad  {\rm and}
\end{equation}
\begin{equation}
{L \over L^* } = \left({\sigma \over 144 km/sec}\right)^{2.6} \quad 
\end{equation}
for early and late type galaxies, respectively, where $L^*$
corresponds to ${M^*}_B = 19.7 + 5 \log h$.  For any assumed
redshift we can calculate a predicted velocity dispersion and distance
modulus,
\begin{equation}
m_{AB}(\lambda_{obs}) - M_{AB}(\lambda_{rest}) 
= 5 \log (D_L) + 7.5 \log(1+z) \quad ,
\end{equation}
where $D_L$ is again an angular diameter distance.  Here $m_{AB}$, the
predicted apparent magnitude, depends both on the measured separation
and the assumed redshift.  Adopting standard absolute spectral energy
distributions (SEDs) for a giant elliptical galaxy and an Scd spiral
from the HST WFPC2 handbook (Burrows {\it et al.} 1995) and normalizing
these at $4000 (1+z)$ \AA \ gives us predicted SEDs.  As redshift increases
the curves move to redder wavelengths and fainter apparent
magnitudes while preserving the shapes of the SEDs.

The photometry of Table 2 can be transformed to the $m_{AB}$ system by
adding -0.17, 0.00, 0.18 and 0.33 mag, respectively, to the observed
$B,~V,~R$ and $I$ apparent magnitudes.  For the sake of comparison we have
computed the predicted SEDs for a giant elliptical at redshift of 0.2
and an Scd spiral at a redshift of 0.4.  These predict a $V$ magnitude
close to the observed value, but it is clear from Figure 5, first,
that the observed colors are too blue and second, that only a very
late type galaxy at very low redshift could produce colors as blue as
those observed.  Such a galaxy would be far brighter than the
third object.  

\subsection{The Third-Image Hypothesis}

As is evident from the above discussion, models interpreting the third
object as the lensing galaxy have serious deficiencies.  Since it is
substantially fainter than the other two, one might suppose the third
object to be a demagnified image associated with the core of the
lensing galaxy.  But unless the core radius were very large, such an
image would lie very close to the center of the galaxy, again giving
an implausibly small impact parameter and the corresponding large
magnifications.  Models of this sort have three additional free
parameters beyond those described in the previous section, a core
radius and the two coordinates of the lensing galaxy (now taken to be
unknown) giving a total of eight free parameters.  The 6 coordinates
for the three components and the two magnification ratios give eight
constraints.  In our (admittedly limited) experimentation with such
models we found the solutions to be quite unstable.

A more exotic variant of the third-image hypothesis is that we are seeing
a three-image configuration associated with a galactic disk (Keeton
and Kochanek 1997).  The disk (like a uniform edge-on sheet) produces
two images which straddle it.  If the disk is not quite edge-on, and
has a finite projected density at its midplane, a third image will
form close to the midplane, in the same way that a third image forms
inside the core radius of a non-singular isothermal sphere.  In the present
case the non-colinearity of the three components would then call for a
degree of central concentration deflecting the midplane image more
than the two straddling images.  Again the added degrees of freedom
make fitting such models a frustrating exercise.

But just as the colors are troublesome when we force object C to have the
same extended shape in all filters, the colors are troublesome when we
assume that object C is a point source.  As described in \S2.3, when we
treat the system as three point sources we find that component C is much
redder than the other two.  This might still be consistent with the
third-image hypothesis, but only if the third component were reddened by
dust in the lensing galaxy.

\subsection{A Working Hypothesis: Blame the Data}

The difficulties with both hypotheses regarding the third object are
traceable to the $B$ filter image.  The $B$ filter position of the
third image forces very round potentials and large magnifications.
The $B$ filter apparent magnitude is likewise the most inconsistent
with either hypothesis.  Since we have only one $B$ exposure in good
seeing, one might imagine some unmodeled phenomenon is affecting the
data -- a cosmic ray or a non-linear pixel response -- or simply an
error in the data reduction.  In many ways this is more palatable than
either physical model for the observations.  Still another alternative
is that the system is yet more complicated with perhaps both a third
quasar component and a lensing galaxy of comparable brightness.

\section{SUMMARY AND CONCLUSIONS}

We have discovered a new gravitationally lensed quasar, with two
components which are confirmed images of the quasar and a third object
which might either be the lensing galaxy or another quasar image.
Both hypotheses for the third object have serious shortcomings.
Possible explanations for these difficulties include corrupt data, an
error in the data reduction and analysis, or a yet more complicated
geometry on scales too small to be resolved with the present
observations.

\acknowledgements

We gratefully acknowledge support for the FIRST survey from NRAO,
the National Science Foundation (grants AST94-19906 and AST94-21178), 
the IGPP/LLNL (DOE contract W-7405-ENG-48), 
the STScI,
the National Geographic Society (grant NGS 5393-094),
NATO (grant CRG 950765),
Sun Microsystems, and 
Columbia University.
PLS gratefully acknowledges financial support from NSF grant
AST96-16866.

\clearpage

\figcaption[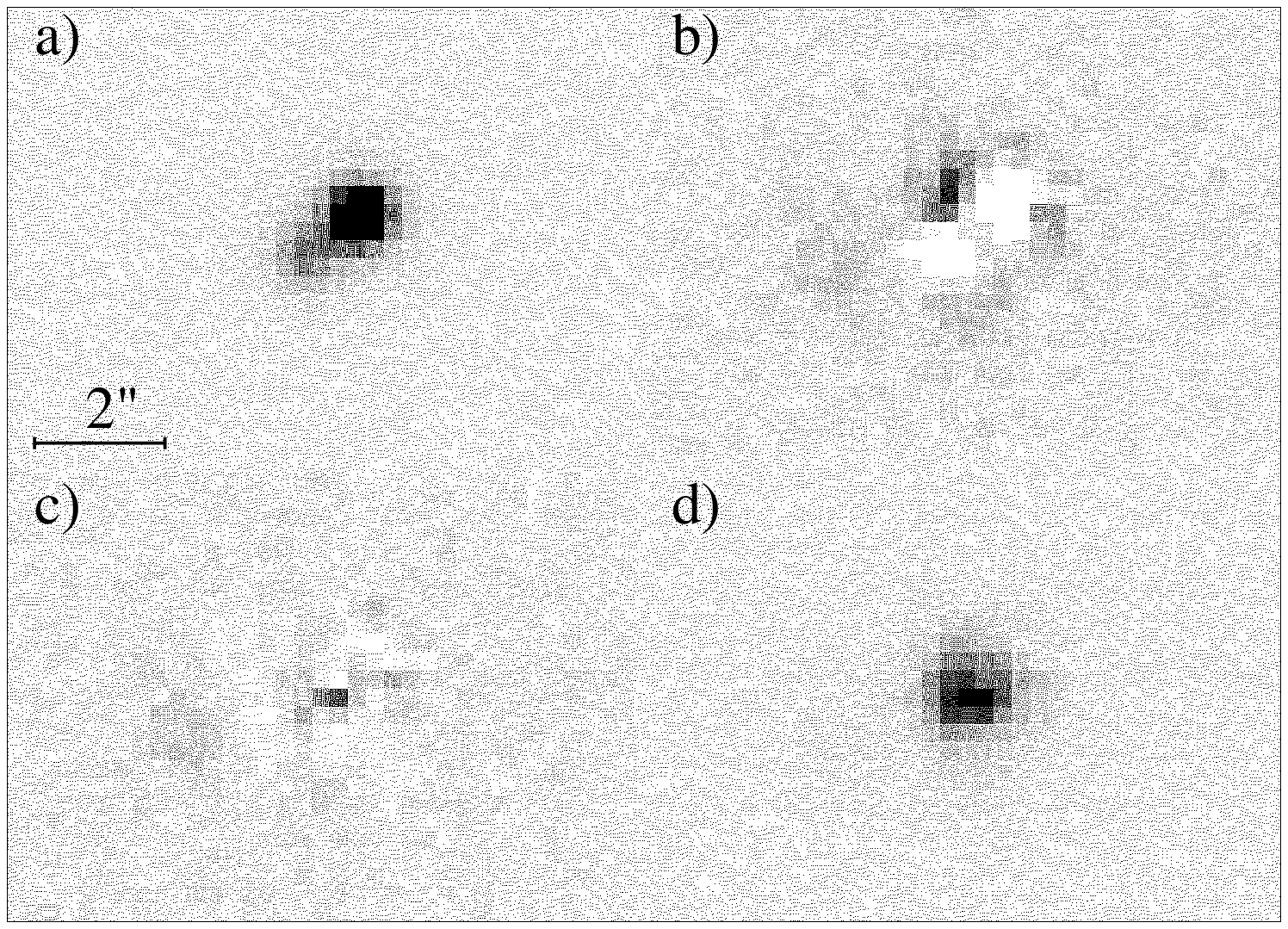]
{ a) A 600 s  R filter direct image of FBQ 0951+2635
at low contrast.  The brighter component, A, is N (up) and W (to the right)
of B.  b) The same image (at higher contrast) after
simultaneous fitting of a stellar PSF to A and B and subtracting.  c)
The same image after simultaneous fitting of two stellar PSFs and a
pseudogaussian and subtracting.  d) The same image with stellar
PSFs subtracted but without subtracting the pseudogaussian.
Panels b), c) and d) are at factors of 50, 50, and 17 higher contrast,
respectively, than panel a).
\label{fig1}}

\figcaption[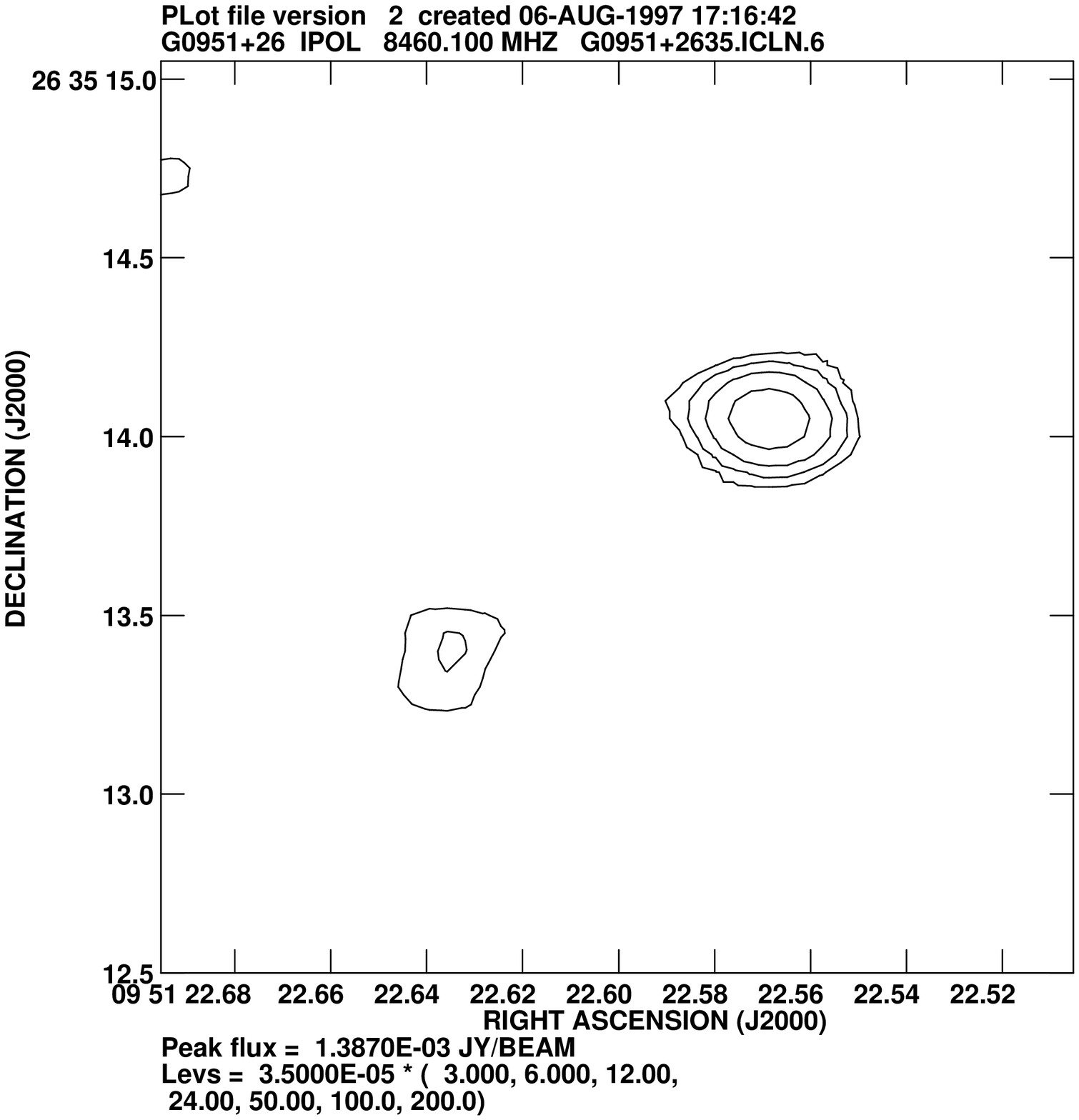]
{VLA A-array map of FBQ 0951+2635 with contours
plotted at 3, 6, 12, 24 and 50 times the RMS noise.
\label{fig2}} 

\figcaption[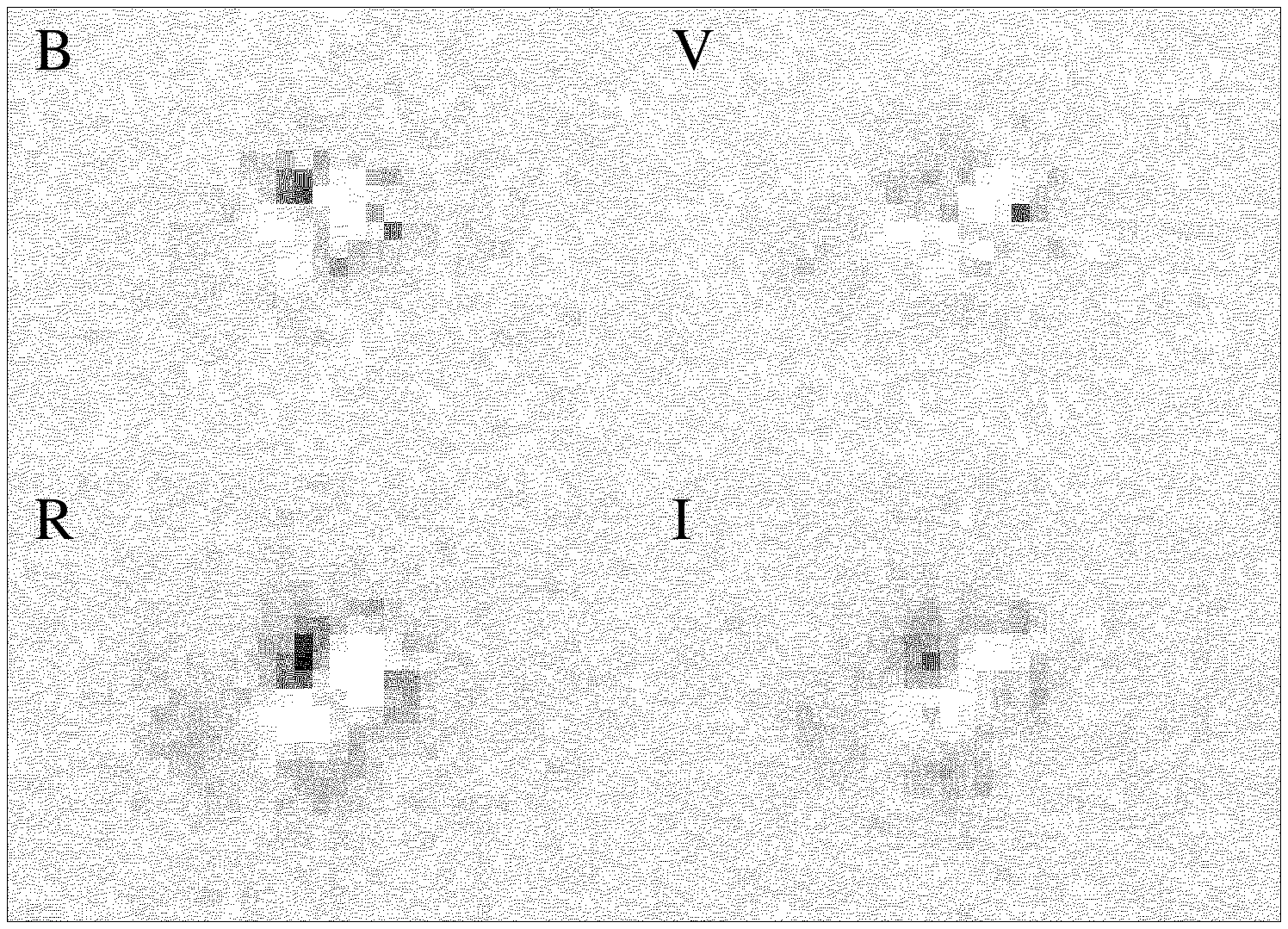]
{$B,~V,~R$, and $I$ images  of FBQ 0951+2635.
North is up and East is to the left.  For each filter empirical
stellar PSFs have been fitted to the A and B components and subtracted
from the image.  The pattern of residuals suggests a third object to
the NE of of A and B.
\label{fig3}}

\figcaption[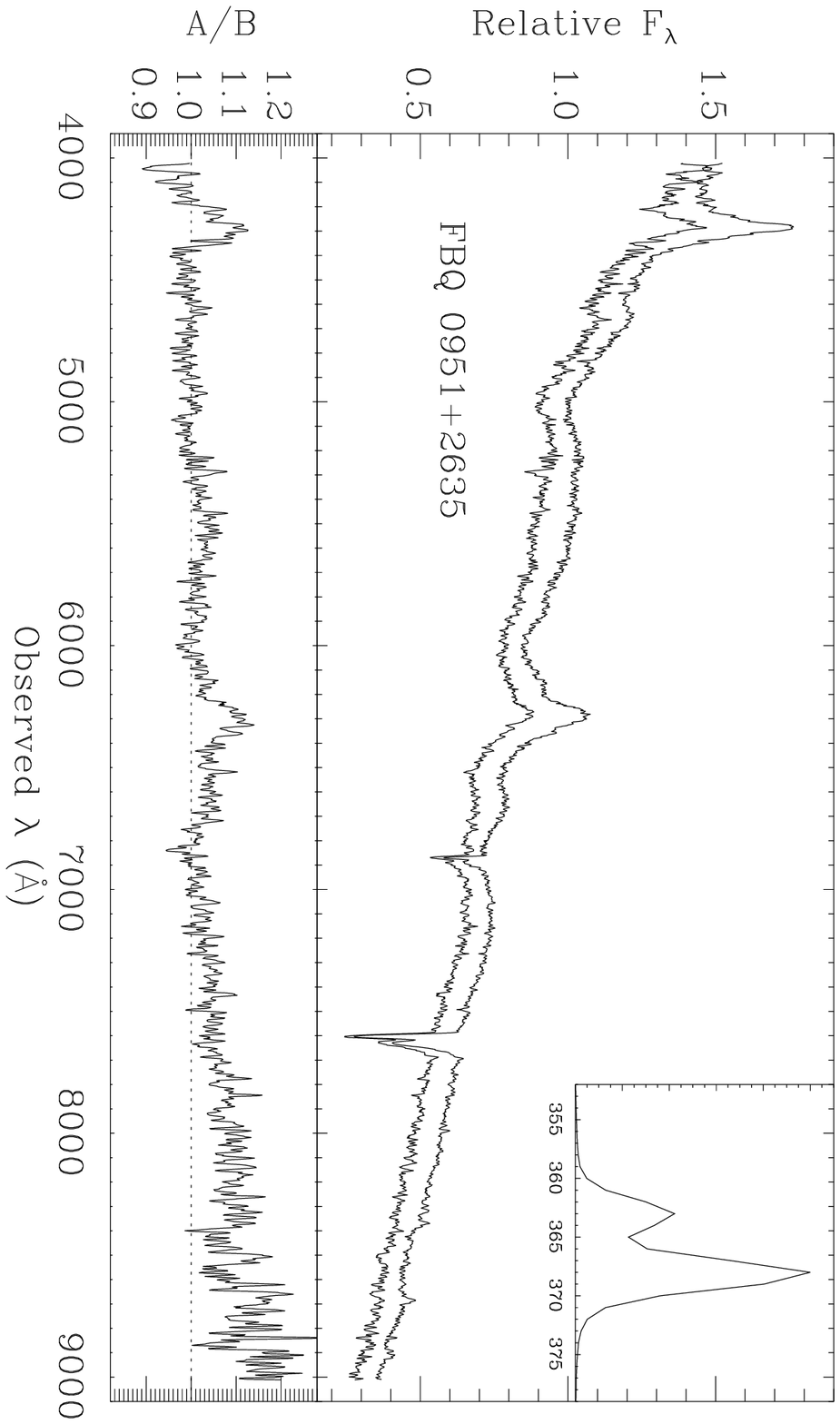]
{Upper panel: Spectra of the A and B components 
of FBQ 0951+2635.  Lower panel: the ratio of the A and B components,
after normalizing at 5000 \AA.  Inset: a cut perpendicular to the
spectrum, showing the 5 pixel separation between the A and B components.
\label{fig4}}

\figcaption[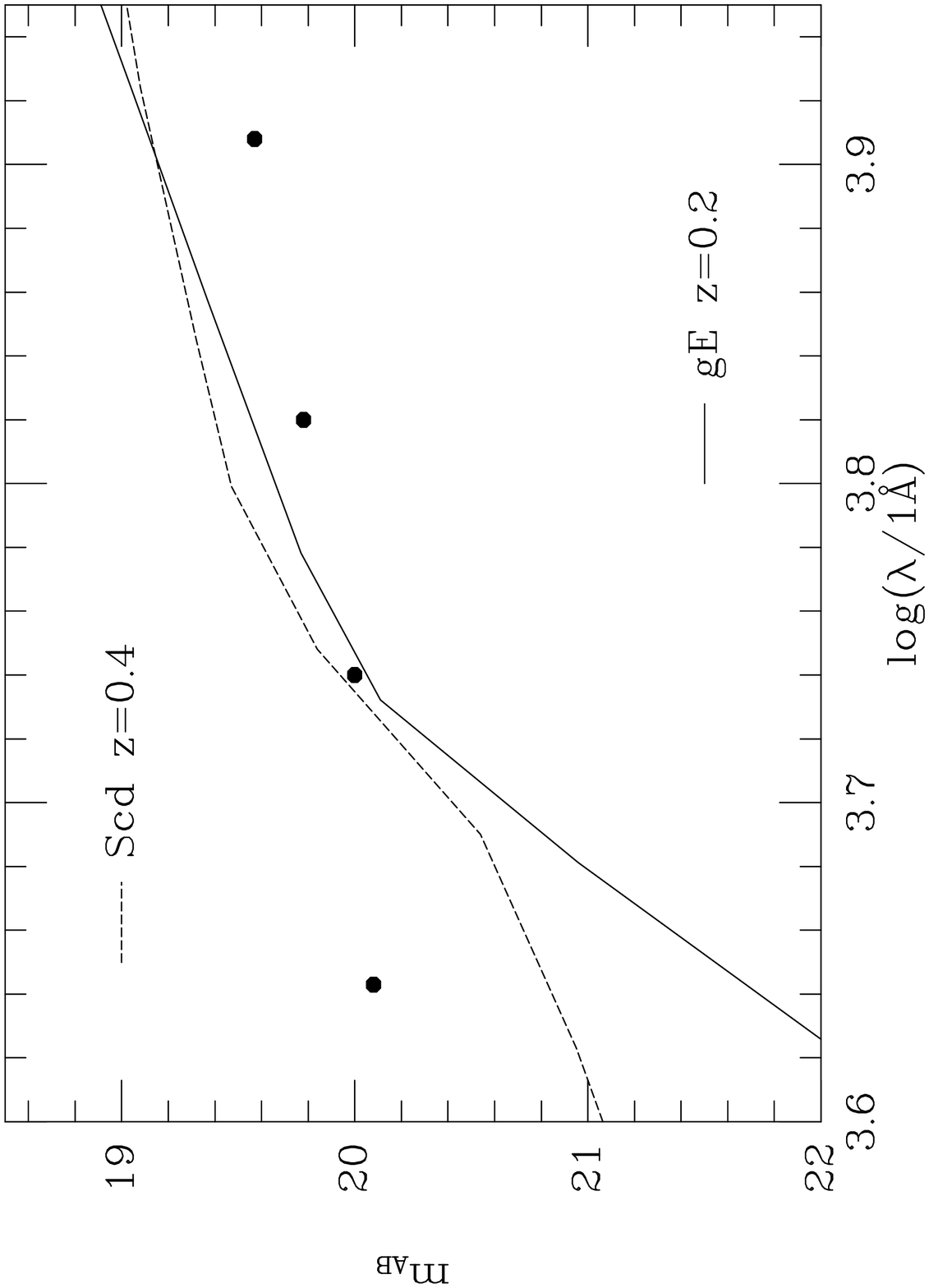]
{Predicted $AB$ apparent magnitude as a function
of wavelength for the lensing galaxy assuming either the SED typical
of a giant elliptical (solid line) galaxy or that of a late type Scd
galaxy (dashed line).  The points are $AB$ magnitudes for the lensing
galaxy corresponding (left to right) to the $B, V, R$, and $I$
photometry of Table 2.  Redshifts for the SEDs were chosen to agree
with the observed image separation and $V$ filter photometry.
\label{fig5}}

\clearpage


\clearpage



 

\makeatletter
\def\jnl@aj{AJ}
\ifx\revtex@jnl\jnl@aj\let\tablebreak=\nl\fi
\makeatother


\begin{deluxetable}{lllr}
\tablewidth{33pc}
\tablecaption{FBQ 0951+2635: VLA A Array Observations}
\tablehead{
\colhead{Obj.}& 
\colhead{RA(2000)}&
\colhead{Dec(2000)}& 
\colhead{$f_{8.4 GHz}$}
}
\startdata
A & 09 51 22.569 & +26 35 14.05 & 1.4 mJy \nl
B & 09 51 22.636 & +26 35 13.38 & 0.3 mJy \nl

\enddata
\end{deluxetable}


\clearpage




 

\makeatletter
\def\jnl@aj{AJ}
\ifx\revtex@jnl\jnl@aj\let\tablebreak=\nl\fi
\makeatother


\begin{deluxetable}{lrrrrrr}
\tablewidth{33pc}
\tablecaption{Astrometry and photometry for FBQ~0951+2635 and nearby stars}
\tablehead{
\colhead{Obj.}& 
\colhead{$\Delta$ RA (s)}&
\colhead{$\Delta$ Dec ('')}& 
\colhead{V}&
\colhead{B$-$V}&
\colhead{V$-$R}&
\colhead{R$-$I}}
\startdata
 A    &0.        &0.               &17.31    &0.35    &0.29   &0.28 \nl

 B   &+0.0669   &-0.638             &18.25    &0.30    &0.27   &0.29 \nl

 C   &+0.0413   &-0.220              &20.00    &0.25    &0.40   &0.36 \nl

05   &+5.678  &-116.27               &16.484   &1.397   &0.878  &0.931 \nl

22   &-1.197   &-36.98               &17.454   &1.332   &0.842  &0.862 \nl

28   &-6.810   &-18.98               &17.119   &0.901   &0.528  &0.513 \nl

30   &-7.472   &+50.88               &16.206   &0.778   &0.451  &0.439 \nl

34   &-8.321   &+32.37               &18.027   &1.602   &1.082  &1.397 \nl

\enddata
\end{deluxetable}


\clearpage

\begin{figure}[h]
\vspace{7.0 truein}
\includegraphics{PaulSchechter.fig1.ps}
\end{figure}

\clearpage

\begin{figure}[h]
\vspace{7.0 truein}
\includegraphics{PaulSchechter.fig2.ps}
\end{figure}

\clearpage

\begin{figure}[h]
\vspace{7.0 truein}
\includegraphics{PaulSchechter.fig3.ps}
\end{figure}

\clearpage

\begin{figure}[h]
\vspace{7.0 truein}
\includegraphics{PaulSchechter.fig4.ps}
\end{figure}

\clearpage

\begin{figure}[h]
\vspace{7.0 truein}
\includegraphics{PaulSchechter.fig5.ps}
\end{figure}

\clearpage

\end{document}